\documentclass[%
 aip,
 jcp,%
 amsmath,amssymb,
 reprint,%
]{revtex4-1}

\usepackage{amsmath}
\usepackage{amssymb}
\usepackage{amsfonts}
\usepackage{graphicx}
\usepackage{subfigure}
\usepackage[version=3]{mhchem} 
\usepackage{multirow}
\usepackage{tabularx}
\usepackage{booktabs}
\usepackage{ctable}
\newcolumntype{Y}{>{\raggedright\arraybackslash}X}      
\newcolumntype{W}{>{\raggedleft\arraybackslash}X}       
\newcolumntype{Z}{>{\centering\arraybackslash}X}        

\usepackage{dcolumn}

\begin{document}

\title{Isobaric first-principles molecular dynamics of liquid water with nonlocal 
van der Waals interactions}

\author{Giacomo Miceli}
\email{giacomo.miceli@epfl.ch}
\affiliation{Chaire de Simulation \`a l'Echelle Atomique (CSEA), %
 Ecole Polytechnique F\'ed\'erale de Lausanne (EPFL), %
 CH-1015 Lausanne, Switzerland}

\author{Stefano de Gironcoli}
\affiliation{Scuola Internazionale Superiore di Studi Avanzati (SISSA) and
DEMOCRITOS Simulation Centre, CNR-IOM, via Bonomea 265, I-34136 Trieste, Italy}

\author{Alfredo Pasquarello}
\affiliation{Chaire de Simulation \`a l'Echelle Atomique (CSEA), %
 Ecole Polytechnique F\'ed\'erale de Lausanne (EPFL), %
 CH-1015 Lausanne, Switzerland}
 
\date{\today}

\begin{abstract}
We investigate the structural properties of liquid water at near ambient conditions
using first-principles molecular dynamics simulations based on a semilocal density 
functional augmented with nonlocal van der Waals interactions.  
The adopted scheme offers the advantage of simulating liquid water at essentially the same 
computational cost of standard semilocal functionals. Applied to the water dimer and 
to ice I$_\textrm{h}$, we find that the hydrogen-bond energy is only slightly 
enhanced compared to a standard semilocal functional. We simulate liquid water
through molecular dynamics in the $NpH$ statistical ensemble allowing for 
fluctuations of the system density. The structure of the liquid departs from 
that found with a semilocal functional leading to more compact structural arrangements. 
This indicates that the directionality of the hydrogen-bond interaction 
has a diminished role as compared to the overall attractions, as expected when 
dispersion interactions are accounted for. This is substantiated through a detailed analysis 
comprising the study of the partial radial distribution functions, various 
local order indices, the hydrogen-bond network, and the selfdiffusion coefficient. 
The explicit treatment of the van der Waals interactions leads to an overall improved 
description of liquid water.
\end{abstract}


\maketitle

\section{Introduction}
\label{sec:intro}
Understanding the structural and dynamical properties of liquid water is of great 
importance due to its ubiquity in biological and chemical processes. 
Nevertheless, giving a comprehensive picture of water and aqueous solutions at 
the molecular level still represents a noticeable challenge.
Molecular dynamics simulations based on empirical force fields have long been the 
technique of choice for investigating the microscopical properties of aqueous 
solutions.\cite{jorgensen_JCP1983,mahoney_JCP2000} However, force field methods 
rely on experimental data and the results depend on how the force field is parametrized.
Furthermore, the parametrization varies from one system to another leading to a 
transferability problem. These limitations are overcome by simulation schemes based 
on first-principles density functionals, in which the energy and the atomic forces 
are evaluated directly from the evolving electronic structure.

With the increasing availability of computer resources and the improvement of 
computational algorithms, first-principles techniques have been extensively 
applied to study the structural and dynamical properties of liquid 
water.\cite{laasonen_JCP1993,sprik_JCP1996,izvekov_JCP2002,%
grossman_JCP2003,vandevondele_JCP2004,sit_JCP2005,lee_JCP2007,schmidt_JPCB2009}
It has become clear that density functionals in which the exchange-correlation energy 
is treated within a generalized gradient approximation lead to overstructured systems
and to underestimated diffusion coefficients. When the description of the 
electronic structure is improved through the use of hybrid density functionals, 
these shortcomings persist.\cite{todorova_JPCB2006,guidon_JCP2008} Similarly,
the explicit consideration of the quantum motion of the nuclei is not sufficient to 
lead to a significant amelioration of the structural properties.\cite{morrone_PRL2008}  
At variance, recent studies have pointed to the critical importance of accounting for  
van der Waals interactions in the description of liquid water.\cite{grimme_JCC2004,%
dion_PRL2004,*dion_ePRL2005,thonhauser_PRB2007,lee_PRB2010,cooper_PRB2010,%
vydrov_JCP2010}

Sophisticated methods such as coupled clusters or M{\o}ller-Plesset perturbation theory
accurately account for van der Waals interactions and have recently been applied to
liquid water.\cite{delben_JPCL2013} However, these methods are still   
computationally too demanding for a widespread use in routine simulations.
More recently, several strategies for including van der Waals interactions 
within density functional simulation schemes have been proposed leading to an 
improved description of the structural
and dynamical properties of liquid 
water.\cite{lin_JPCB2009,schmidt_JPCB2009,wang_JCP2011,linJCTC2012}
In particular, Schmidt \textit{et al.} performed molecular dynamics simulations of 
liquid water in the isobaric-isothermal ensemble ($NpT$).\cite{schmidt_JPCB2009} 
In $NpT$ simulations, the supercell parameters are free to adjust during the evolution 
allowing the system to reach its equilibrium density at a fixed pressure 
without imposing additional constraints. These simulations\cite{schmidt_JPCB2009}
revealed that the equilibrium density yielded by semilocal density functionals 
underestimates the experimental value by about 15\%\ and that the experimental density
is recovered through the use of empirical dispersion corrections.\cite{grimme_JCC2004}   
These results are in perfect agreement with the molecular dynamics simulations 
performed by Wang \textit{et al.},\cite{wang_JCP2011} who compared semilocal 
and van der Waals corrected density functionals using a $NVT$ protocol.

Recently, several nonlocal formulations have been introduced which explicitly account 
for van der Waals interactions through a functional of the density.%
\cite{dion_PRL2004,tkatchenko_PRL2009,lee_PRB2010,vydrov_JCP2010}
Among these, the one proposed by Vydrov and Van Voorhis\cite{vydrov_JCP2010} 
in its revised form denoted rVV10 carries the advantage of being particularly 
suited to be used in conjunction with plane-wave basis sets, leading to a scheme 
which does not yield any significant
overhead with respect to standard semilocal functionals.\cite{sabatini_PRB2013}
While rVV10 still relies on a phenomenological formulation, it carries an important 
advantage with respect to empirical schemes. Indeed, once the parameters are set, 
the resulting functional can be carried over to any weakly bonded system without 
requiring any further parameter tuning. It could thus be envisaged to use
the same theoretical scheme for studying various aqueous solutions.

In this work, we present results obtained through first-principles molecular 
dynamics simulations in the $NpH$ ensemble of liquid water at near ambient conditions.
The van der Waals interactions are explicitly taken into account through the functional  
rVV10.\cite{vydrov_JCP2010,sabatini_PRB2013}  
First, we test the performance of the rVV10 functional on the water dimer and
on the I$_\textrm{h}$ phase of ice, finding only small variations of 
the hydrogen bond strength in those systems compared to semilocal functionals.  
In our molecular dynamics simulations of liquid water, we then find that rVV10 
provides a noticeably improved description compared to standard semilocal 
functionals. Not only the equilibrium density is recovered but also the structural 
and dynamical properties are much closer to experimental observations. 
In particular, we focus on the local order and on an analysis of the hydrogen 
bond network.  Hence, rVV10 allows for a good description of liquid water 
at a computational cost equivalent to semilocal density functionals and 
appears particularly attractive for the treatment of water-solute interactions.

The present paper is organized as follows. In Sec.\ \ref{sec:methods},
we describe the methods used in this work and give a detailed account
of our simulation protocols. In Secs.\ \ref{sec:dimer} and \ref{sec:ice},
we apply rVV10 to the water dimer and the I$_\textrm{h}$ phase of ice, 
respectively. The results for liquid water and the corresponding analyses 
are given in Sec.\ \ref{sec:water}. The conclusions are drawn in 
Sec.\ \ref{sec:conclusions}.

\section{Methods}
\label{sec:methods}

\subsection{Computational details}
\label{subsec:details}
The computational scheme adopted throughout this work is based on the self-consistent 
Kohn-Sham approach to density functional theory (DFT) as implemented in the
Quantum-\textsc{espresso} suite of programs.\cite{quantum_espresso}
The valence Kohn-Sham orbitals are expanded up to a kinetic-energy cutoff of 85 Ry, while
core-valence interactions are described by norm-conserving pseudopotentials.\cite{troullier_PRB1991}
The exchange-correlation (XC) energy is described within the generalized-gradient approximation 
proposed by Perdew, Becke, and Ernzerhof (PBE).\cite{pbe} In addition to the semilocal PBE 
functional, we performed calculations in which van der Waals interactions are explicitly 
taken into account. To this aim, among the several van der Waals density functional 
schemes proposed, we adopted rVV10 functional, recently introduced by
Sabatini {\it et al.}\cite{sabatini_PRB2013} as an essentially equivalent variant of the 
nonlocal VV10 functional developed by Vydrov and Van Voorhis.\cite{vydrov_JCP2010}  
The VV10 functional has been shown to be particularly successful in reproducing the  
physical properties of molecules and weakly bonded solids.\cite{vydrov_JCP2010,hujo_JCTC2011} 
The advantage of adopting the revised functional rVV10 rests in a more efficient 
evaluation of the correlation energy and its derivatives within a plane-waves framework, 
a major prerequisite when long \textit{ab initio} molecular dynamics simulations 
have to be carried out.

The rVV10 functional is defined by a very simple analytic form and depends on an 
empirically determined parameter, $b$, which controls the short-range behavior of 
the functional.\cite{vydrov_JCP2010,sabatini_PRB2013} 
This parameter can be optimized in order to reproduce the correct physical properties 
of a chosen set of materials. In particular, the rVV10 functional, as implemented in the 
Quantum-\textsc{espresso} package, has been tested over the S22 set of molecules. 
The best description of the binding energies is obtained for $b=6.3$.\cite{sabatini_PRB2013}
Although accurate in describing the binding energy and geometry of 
molecules,\cite{vydrov_JCP2010,hujo_JCTC2011} the VV10 functional gives overestimated 
binding energies for weakly bonded solids.\cite{bjorkman_PRL2012} 
To achieve an improved description of a set of layered solids, Bj\"orkman has proposed 
a higher value for the parameter $b$ (up to 10.25).\cite{bjorkman_PRB2012} 
To identify the optimal functional for treating aqueous systems, we therefore 
consider in the following various values of the parameter $b$.


\subsection{Molecular dynamics simulations}
\label{subsec:md}

We perform first-principles Born-Oppenheimer molecular dynamics  
simulations of 64 water molecules in the isobaric-isoenthalpic statistical ensemble
($NpH$) at near-ambient conditions. The quantum motion of the nuclei is neglected 
in the present work. We use the Beeman algorithm to integrate Newton's equations
of motion, as implemented in the Quantum-\textsc{espresso} package, with an 
integration time step of 0.48 fs.
Simulations are carried out using the PBE and rVV10 functionals. In the latter,
the parameter $b$ is set to 6.3, 8.9, 9.3, and 9.5, respectively. In 
the following, we use the notation rVV10-b6.3 to indicate the rVV10 functional
with the parameter $b$ set to 6.3, and similarly for the other values of $b$. 
The molecular dynamics runs last between 25 and 35 ps. The cell size is 
allowed to fluctuate with the constraint of conserving the initial cubic 
symmetry. A vanishing external pressure is imposed through the use of the 
Parrinello-Rahman barostat.\cite{parrinello_PRL1980}

The starting configuration for each simulation is chosen from a trajectory obtained 
through a $NVT$ classical molecular-dynamics simulation in which the density 
of the system is fixed at 1 g/cm$^3$ and the temperature at 300 K. For each functional,
a preliminary $NpT$ equilibration run of about 5~ps is carried out before collecting statistics.  
The temperature is set to 350 K to ensure a liquid-like behavior  
\cite{vandevondele_JCP2004,sit_JCP2005,schwegler_PRL2000} During the equilibration runs the 
temperature is controlled by a velocity rescaling thermostat, which only acts
when the cumulative average temperature moves out of the window 350$\pm$20 K.
However, we observe that the temperature drift during the equilibration time is
minimal and that the thermostat remains inactive after an initial period of a 
few picoseconds. In particular, we obtained average temperatures of 349, 344,
and 342 K for our molecular dynamics simulations based on PBE, rVV10-b6.3,
rVV10-b9.3, respectively.

In the $NpT$ molecular dynamics simulations with a basis set 
involving a constant number of plane waves, fluctuations of the volume imply 
fluctuations of the effective energy cutoff defining the basis set.\cite{bernasconi1995}
This condition requires extra precautions. All the simulations are  
started from a geometry with a cell of fixed density at 1 g/cm$^3$ and 
the corresponding cut-off energy is chosen in such a way that the stress 
tensor is converged within 1 kbar.
Those simulations, which show a decrease of density during the equilibration
run, are restarted with a larger initial volume to restore the originally set 
higher cut-off. Once the cumulative average value of the density reaches a constant
value, we ensure that the density fluctuations do not lead to any integration problem.

During all the production runs, we observe that the temperature remains constant 
and that the conditions to activate the thermostat are not met. In these conditions
our simulations are effectively sampling the isobaric-isoenthalpic ($NpH$) ensemble. 
Since the external pressure is forced to be zero the total energy of the system, $E$, 
is conserved on average and can be used to check the quality of our integration scheme.

\section{Water dimer}
\label{sec:dimer}

The simplest system that water molecules can form is the water dimer. 
Despite its simplicity, the water dimer has extensively been investigated 
in physical chemistry.\cite{dyke_JCP1977,odutola_JCP1980}
It plays a major role in many important physical and chemical processes. 
In addition, a better understanding of the isolated water dimer implies 
a better understanding of the hydrogen bond which is at the origin
of the very peculiar properties of water in its condensed phases.\cite{jeffrey1997introduction}
Finally, in simulation studies, the water dimer is often used as 
the simplest water system to test new computational approaches.\cite{kim_JPC1994,feyereisen_JPC1996}

\begin{figure}
\centering
\includegraphics[width=8.5cm]{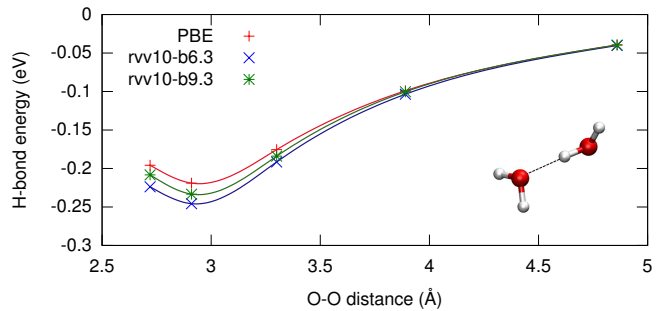}%
\caption{Binding energy of the water dimer as a function of the intermolecular 
         separation calculated with the PBE, rVV10-b6.3, and rVV10-b9.3 functionals.
         The water dimer geometry is shown in the inset.}
 	\label{fig:benergy}
\end{figure}

In Fig.\ \ref{fig:benergy}, we report the binding energy of the water dimer as
a function of the inter-molecular separation, as calculated with various 
functionals. The binding energies are obtained for a dimer with fixed 
geometries in a cubic supercell with lattice parameter of 20 \AA. 
We took as our reference structure the optimized geometry obtained in 
Ref.\ \citenum{jurevcka_PCCP2006} through the coupled cluster CCSD(T) 
method in the complete basis set limit.
The structures with other O-O distances, were then obtained by displacing
one of the H$_2$O molecules while preserving the symmetry and keeping the
the angle along the H-bond, $\angle$O$\cdots$HO, fixed and equal to its value
in the reference configuration, $\alpha = 172.8^{\circ}$.
The other structural parameters were not varied with respect to the reference 
structure. The reference dimer geometry is shown in the inset of 
Fig.\ \ref{fig:benergy}.

\begin{table}
 \caption{Binding energy of the water dimer calculated with the PBE, rVV10-b6.3, and
          rVV10-b9.3 functionals.}
 \label{tab:dimer}
  \begin{ruledtabular}
  \begin{tabular}{lc}
   & $E_0$/H$_2$O (eV) \\
  \hline
   PBE         & 0.22 \\
   rVV10-b6.3  & 0.25 \\
   rVV10-b9.3  & 0.23 \\
   CCSD(T)\cite{jurevcka_PCCP2006} & 0.22 \\
  \end{tabular}
 \end{ruledtabular}
 
\end{table}

The optimized CCSD(T) configuration gives the minimum binding energy for all our
DFT calculations at the same geometry. The obtained values are given in Table \ref{tab:dimer}.
Among the DFT calculations, PBE gives the best agreement with our reference 
calculation.\cite{jurevcka_PCCP2006} However, the binding energies obtained 
with the various VV10 variants increase by at most 0.03 eV.
For each of the DFT functionals, we check that a full structural relaxation
leads to energy differences of less than 5 meV without causing any 
significant structural modification with respect to the reference configuration.  
These results indicate that all the considered DFT functionals account well for the 
strength of the H-bond and for the geometry of the water dimer.

\section{I$_\textrm{h}$ phase of ice}
\label{sec:ice}
We next investigate the effect of including van der Waals interactions on 
the H-bond strength, cohesive energy, and structural properties of ice. 
The performance of the semilocal PBE and the various rVV10 functionals
are compared among each other and against experimental and theoretical data.

\begin{figure}
\centering
\includegraphics[width=6.5cm]{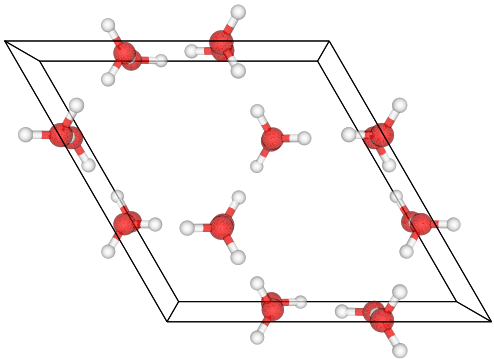}%
\caption{Perspective view along the $c$ axis of the Bernal-Fowler model of ice I$_\textrm{h}$.}
 	\label{fig:icestructure}
\end{figure}

At present, several ice phases have been experimentally characterized. The 
oxygen atoms are always found in an ordered sublattice and the water molecules are 
orientated in such a way that the O atoms are approximately tetrahedrally coordinated
by four H atoms, two of which are bonded covalently while the other two belong 
to nearest-neighbor water molecules and are bonded through hydrogen bonds.
This ``ice rule'' is always satisfied even if the varying orientation of the 
water molecules may lead to disordered networks lacking any form of translational 
invariance. The various ice structures therefore mainly differ by the form of the 
hydrogen-bond network and the packing density. 

For our purpose, we use the periodic model of the I$_\textrm{h}$ phase proposed 
by Bernal and Fowler.\cite{bernal_JCP1933} 
This model of ice has been adopted in several theoretical studies to test 
the accuracy of density functionals in describing solid water 
networks.\cite{hamann_PRB1997,brian_PRB2011,santra_JCP2013}
Figure \ref{fig:icestructure} shows a top-view perspective along the $c$ axis of 
the ordered Bernal-Fowler model of ice. The unit cell contains 12 water molecules
and belongs to the P6$_3$cm(C$^3_{6v}$) space group. 
Following Hamann,\cite{hamann_PRB1997} we adopt the experimental cell ratio
$c/a$, which ensures that all hydrogen bonds have the same length.
The Brillouin-zone is sampled using a 2$\times$2$\times$2 Monkhorst-Pack grid 
which does not include the $\Gamma$-point. 

\begin{figure}
\centering
\includegraphics[width=8.5cm]{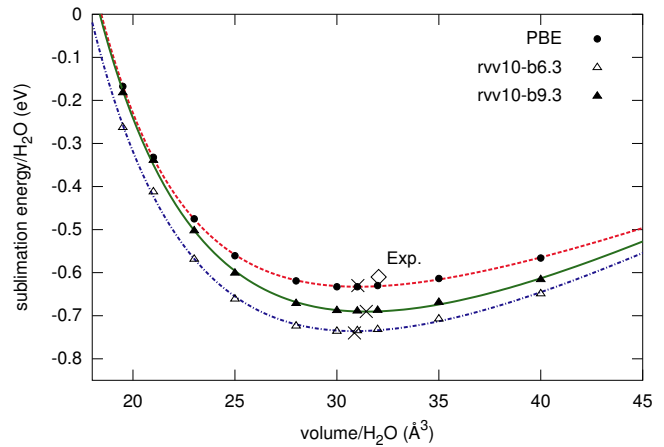}%
\caption{Sublimation energies per H$_2$O molecule vs.\ volume for ice I$_\textrm{h}$ 
         as obtained with PBE, rVV10-b6.3, rVV10-b9.3 functionals. The minima of the Murnaghan
         equation of state are marked and the experimental value is reported for
         comparison.\cite{whalley_JCP1984,brill_ActaC1967}}
 	\label{fig:iceEoS}
\end{figure}

In Fig.\ \ref{fig:iceEoS}, the calculated sublimation energies as a function 
of the volume per water molecule are plotted for PBE, rVV10-b6.3, and rVV10-b9.3 
functionals. In this calculation, the $c/a$ is not allowed to change. 
The energetics and the equilibrium structural properties of ice 
I$_\textrm{h}$ are reported in Table \ref{tab:ice}. The PBE functional gives  
a sublimation energy of 0.63 eV, slightly larger than 
the experimental value of 0.61 eV. This result is in good agreement with previous 
theoretical results which generally overestimate the experimental value 
by an amount ranging between 30 and 100 meV per H$_2$O molecule. In particular, it is 
noteworthy that our results agree closely with the highly converged projector-augmented-wave
calculations reported in Ref.\ \citenum{feibelman_PCCP2008}.
The account of van der Waals interactions leads to an even larger overestimation
of the experimental results. This result qualitatively agrees with a previous 
theoretical investigation of ice I$_\textrm{h}$ by Santra {\it et al.}\cite{santra_JCP2013}
However, as can be seen in Table \ref{tab:ice}, a general improvement is achieved 
when the phenomenological parameter $b$ in the rVV10 formulation 
is set to 9.3 rather than to its original value of 6.3.

\begin{table}
\caption{Sublimation energy per H$_2$O molecule $E_0$, equilibrium volume per H$_2$O 
          molecule $V_0$, 
          and bulk modulus $B$ of ice I$_\textrm{h}$ as obtained with PBE, rVV10-b6.3, 
          and rVV10-b9.3 functionals.  } 
 \label{tab:ice}
 \begin{ruledtabular}
  \begin{tabular}{lccc}
   & $E_0$ (eV) &%
   $V_0$ (\AA$^3$) &%
   $B$ (GPa) \\
   \hline
   PBE         & 0.63  & 31.03 & 13.2  \\
   rVV10-b6.3  & 0.74  & 30.86 & 16.2  \\
   rVV10-b9.3  & 0.69  & 31.44 & 15.6  \\
   Expt.\cite{whalley_JCP1984,brill_ActaC1967,hobbs1974} & 0.61 & 32.05 & 10.9 \\
  \end{tabular}
 \end{ruledtabular}
\end{table}

\section{Liquid water}
\label{sec:water}
We here present our results obtained through molecular dynamics
simulations of liquid water near-ambient conditions using both semilocal 
and van der Waals corrected density functionals. All the data presented here 
are the result of statistical analyses performed on molecular dynamics 
trajectories having duration between 25 and 35 ps.
For technical details of the simulations we refer the reader to Sec.\ \ref{subsec:md}.

\subsection{Equilibrium density}
\label{subsec:density}

\begin{figure}
\centering
\includegraphics[width=8.5cm]{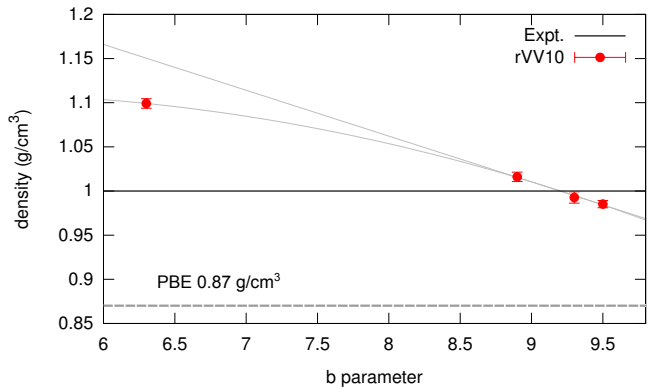}%
\caption{Equilibrium density of liquid water at near-ambient condition as obtained
         from rVV10 simulations with different values of the empirical $b$ parameter.
         For comparison, the equilibrium densities obtained in the experiment and
         in simulations based on the PBE functional are also reported.
         All the simulations
         show average temperatures falling within the range $345\pm 4$ K. 
         The solid horizontal line indicates the density of liquid water at 
         4$^\circ$C (1 g/cm$^3$).  }
 	\label{fig:density}
\end{figure}

Isobaric molecular dynamics simulations allow for a direct evaluation
of the equilibrium density of the system. We find that the 
PBE functional gives an average density for liquid water of $\sim$0.87 g/cm$^3$ 
underestimating the experimental value by about 13\%, in very good agreement with 
previous theoretical studies.\cite{schmidt_JPCB2009,wang_JCP2011}
In contrast, the explicit introduction of van der Waals interactions, described 
by the rVV10 functional with the  parameter $b$ set to its original value of 6.3, causes
an overestimation of the equilibrium density. We obtain a value of $\sim$1.10 g/cm$^3$, 
about 10\%\ higher than the experimental value. 
Figure \ref{fig:density} shows the average densities obtained with rVV10 simulations 
for various values of the parameter $b$. Our simulations give average densities 
of 1.10, 1.02, 0.99, and 0.98 g/cm$^3$ for $b$=6.3, 8.9, 9.3, and 9.5, respectively.
In particular, for the specific value of $b$=9.3 the simulation yields an average 
density which closely matches the experimental density of 1 g/cm$^3$. 

While a higher density clearly corresponds to more compact arrangements and to a 
partial breakdown of the hydrogen-bond network, this result does not come from a
weakening of the hydrogen-bond strength. In Sec.\ \ref{sec:dimer}, we indeed show
that the rVV10 functional leads to a slight increase of the hydrogen-bond energy
when compared to the PBE one. Therefore, the more compact arrangements found with
rVV10 necessarily come from alternative structural configurations which are more
favored than in the PBE, due to the less directional nature of the attraction between
the water molecules.

\begin{figure}
\centering
\includegraphics[width=8.5cm]{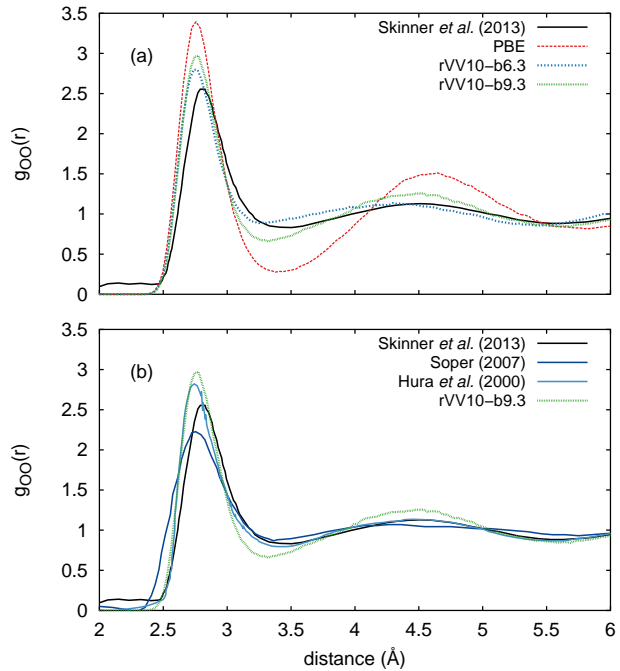}%
\caption{(a) The oxygen-oxygen radial distribution functions 
         $g_{\textrm{OO}}(r)$ as obtained with PBE, rVV10-b6.3, and rVV10-b9.3 
         functionals are compared with the x-ray diffraction result of Skinner 
         {\em et al.}\ from Ref.\ \onlinecite{skinner_JCP2013}. 
         (b) The $g_{\textrm{OO}}(r)$ as obtained with rVV10-b9.3 is 
         compared with various experimental measurements.\cite{skinner_JCP2013,%
         soper_JPCM2007,hura_JCP2000} The average temperatures are 349, 344, 
         342 K for PBE, rVV10-b6.3, and b-9.3 simulations, respectively, while 
         all the experimental results have been obtained at ambient temperature.}
 	\label{fig:gor}
\end{figure}

\begin{figure}
\centering
\includegraphics[width=8.5cm]{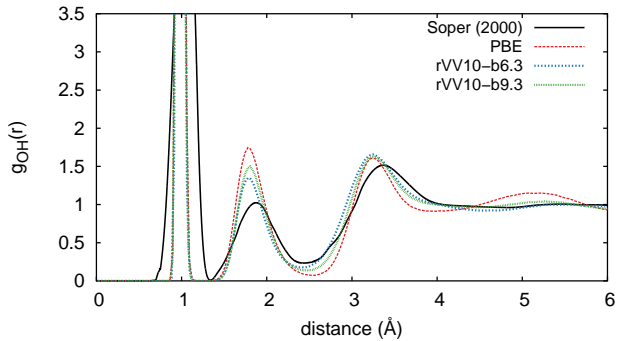}%
\caption{Oxygen-hydrogen radial distribution functions as obtained with
         PBE, rVV10-b6.3, and rVV10-b9.3 functionals are compared with the experimental
         result from Ref.\ \citenum{soper_CP2000}. The average temperatures in 
         the simulations are reported in the caption of Fig.\ \ref{fig:gor}.}
 	\label{fig:gor-bis}
\end{figure}

\subsection{Radial distribution functions}

The oxygen-oxygen and oxygen-hydrogen radial distribution functions (RDF) 
as obtained with various functionals are compared with recent experimental 
results\cite{skinner_JCP2013,soper_JPCM2007,hura_JCP2000,soper_CP2000}
in Figs.\ \ref{fig:gor} and \ref{fig:gor-bis}, respectively. 
For the radial distribution functions achieved with the PBE functional,  
one notices that the obtained description of liquid water is overstructured, 
in agreement with previous theoretical studies.\cite{schmidt_JPCB2009}
Focusing on O-O correlations, Fig.\ \ref{fig:gor}(a), we remark that the position
of the first peak at 2.75 \AA\ in the simulated radial distribution function is 
in good agreement with the experimental position at 2.80 \AA, but the second 
coordination shell is slightly shifted towards larger distances, and so 
is the second minimum.

An overall improved description is achieved for the rVV10 functionals.
When the rVV10 functional is used in its original form with $b=6.3$, 
the agreement with the experimental result improves as far as the 
value of $g_{\textrm{OO}}^{\textrm{max}}$ is concerned. 
An improvement is also observed for the second solvation shell, where 
the $g_\textrm{OO}(r)$ is nevertheless less structured with respect to the experiment   
and the position of the peak is slightly shifted toward lower distances.
The rVV10 functional with $b=9.3$ yields yet better positions of minima and 
maxima, but the $g_\textrm{OO}(r)$ is slightly overstructured.
In Fig.\ \ref{fig:gor}(b), we compare the $g_{\textrm{OO}}(r)$
obtained with the rVV10-b9.3 functional with other experimental 
results in literature.\cite{skinner_JCP2013,soper_JPCM2007,hura_JCP2000}
This comparison shows the variation among the experimental results, but 
confirms that the $g_{\textrm{OO}}(r)$ obtained with the rVV10-b9.3 
functional is slightly overstructured. 

For oxygen-hydrogen correlations (Fig.\ \ref{fig:gor-bis}), the rVV10 
functionals similarly yield smoother radial distribution functions 
than the PBE, but the dependence on the parameter $b$ 
appears less pronounced. The closer agreement with experiment for the radial 
distribution functions is consistent with the improved description of  
the equilibrium density (cf.\ Sec.\ \ref{subsec:density}). 

The agreement with experiment achieved with the functional rVV10 is overall similar 
to that obtained with empirical van der Waals interactions.\cite{schmidt_JPCB2009}
Simulations based on the rVV10 functional with values of the $b$ parameter ranging 
from 8.9 to 9.5 show structural properties with minimal differences. Therefore,
we focus in the following only on results obtained with PBE, the original rVV10-b6.3, 
and rVV10-b9.3. 

\subsection{Local order}

An analogy can be drawn between variations of the parameter $b$ in the rVV10 functional 
and variations of the external pressure. As shown in Fig.\ \ref{fig:density},
the equilibrium density of liquid water ranges from 1.10 to 0.87 g/cm$^3$ when going 
from a rVV10 simulation with the lowest value of the parameter $b$ to PBE. 
Accordingly, the liquid-water system undergoes an increase of the local order 
as can be inferred from the oxygen-oxygen radial distribution functions in Fig.\ \ref{fig:gor}.
In order to give a microscopic interpretation of the density variations resulting 
from the tuning of the phenomenological parameter $b$, we perform statistical 
analyses of structural descriptors which give a measure of the local order.

\begin{figure}
\centering
\includegraphics[width=7.5cm]{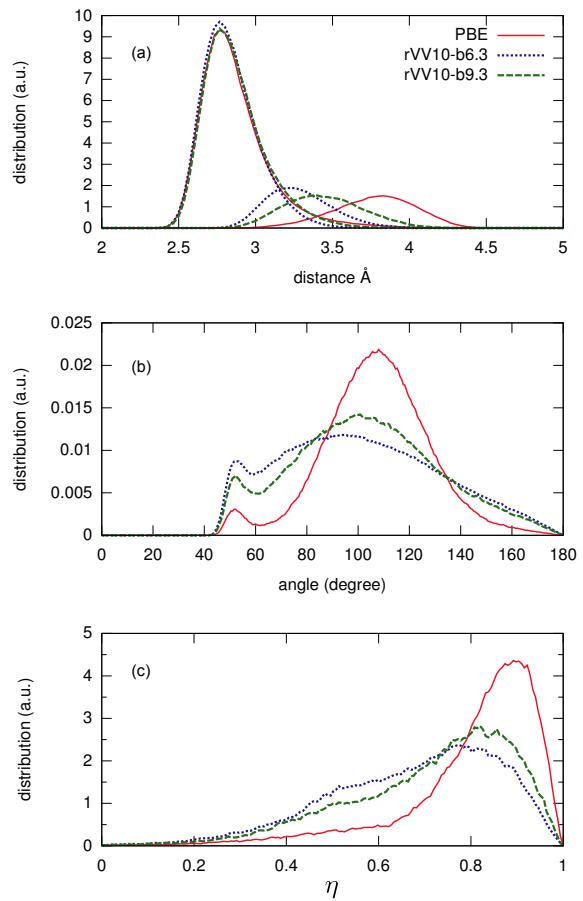}%
\caption{Distributions of local order parameters obtained with PBE, rVV10-b6.3,
         and rVV10-b9.3 functionals: 
         (a) O-O distance within the first coordination shell, 
             the first four neighbors and the fifth one are 
             presented separately; 
         (b) $\angle$ O-O$_\textrm{c}$-O angle for 
             an O-O$_\textrm{c}$ cutoff distance of 3.5 \AA;
         (c) orientational order parameter $\eta$ defined 
             in Eq.\ \ref{eq:qparam}.}
 	\label{fig:distance}
\end{figure}

We perform an analysis of the O-O distances between a water molecule and other water 
molecules belonging to its first solvation shell. For each instantaneous solvation shell
around a central molecule, the neighboring molecules are ranked according to theses distances.
The first four molecules are assumed to form the first coordination shell. 
As shown in Fig.\ \ref{fig:distance}(a), all the functionals considered in this work
produce very similar distance distributions for the first coordination shell.
At variance, the distribution of the fifth water molecule is highly sensitive to the adopted
functional and moves closer to the central water molecule, as the contribution of the
dispersion forces becomes more important. In the literature, this fifth water molecule is
often interpreted as an ``interstitial'' molecule within the regular liquid water 
network.\cite{saitta_PRE2003}

Additional insight into the local arrangement of water molecules can be acquired by 
inspecting the O-O$_\textrm{c}$-O angle distributions given in Fig.\ \ref{fig:distance}(b), 
where the O$_\textrm{c}$ atom belongs to the central molecule while the other O atoms 
belong to the solvation shell defined by an O-O$_\textrm{c}$ cutoff distance of 3.5 \AA. 
When compared to PBE, the rVV10 functionals favor a shift 
towards lower angles and a broadening of the main peak at about 109$^\circ$.  
This clearly indicates a distortion of the tetrahedral geometry of the first 
coordination shell.

The tetrahedral distortion can be further investigated by the distribution of
the orientational order parameter $\eta$ introduced in Ref.\ \citenum{errington_Nature2001}: 
\begin{equation}
 \eta = 1 - \frac{3}{8} \sum_{i=1}^3 \sum_{j=i+1}^4 \left( \cos \theta_{ij} + \frac{1}{3} \right)^2
 \label{eq:qparam}
\end{equation}
where $\theta_{ij}$ is the angle between the vectors connecting the oxygen atom
of the central molecule with two oxygen atoms belonging to its first four nearest-neighbors 
molecules. This parameter is defined in such a way that it assumes the value $\eta=1$ in a 
perfect tetrahedral geometry such as in ice I$_\textrm{h}$, whereas $\eta=0$ in an ideal gas. 
As shown in Fig.\ \ref{fig:distance}(c), the PBE gives a distribution which peaks at
a high value of $\eta$ around 0.9. When van der Waals interactions are explicitly accounted 
for, the main peak of the distribution shifts to lower values and a secondary feature 
appears around 0.5.  In their original work, Errington and Debenedetti\cite{errington_Nature2001} 
found that the relative weight of the second feature increased with temperature, i.e.\ 
with structural disorder. In our simulations, the enhanced weight of the second feature 
is related to the decrease of the parameter $b$, or in other words with the increase of 
the equilibrium density. 

Another order parameter which is sensitive to the degree of local tetrahedral order is the 
local structure index (LSI) $I$ introduced in Refs.\ \citenum{shiratani_JCP1998,palmer_Nature2014}.
This parameter is expected to discriminate molecules with a well structured tetrahedral environment,
with separated first and second shells, from molecules around which the order is perturbed 
by an approaching ``interstitial'' molecule.\cite{shiratani_JCP1998,palmer_Nature2014}
For a given central molecule $\mu$, a neighboring oxygen atom $i$ is ranked and labelled according 
to the distance $r_i$ to the central molecule $\mu$, such that :
$r_1 < r_2 < \dots < r_i < r_{i+1} < \dots < r_{n(\mu,t)} < r_{n(\mu,t)+1}$, where 
the number $n(\mu,t)$ satisfies the relation $r_{n(\mu,t)} < 3.7\, \textrm{\AA} < r_{n(\mu,t)+1}$.
The LSI $I(\mu,t)$ of the central molecule $\mu$ at time $t$ then reads:
\begin{equation}
 I(\mu,t) = \frac{1}{n(\mu,t)} \sum_{i=1}^{n(\mu,t)} \left[ \Delta(i;\mu,t) - \bar{\Delta}(\mu,t) \right]^2,
\label{eq:lsi}
\end{equation}
where $\Delta(i;\mu,t) = r_{i+1} - r_{i}$ and $\bar{\Delta}(\mu,t)$ is the average obtained as
\begin{equation} 
\bar{\Delta}(\mu,t) = \frac{1}{n(\mu,t)} \sum_{i=1}^{n(\mu,t)} \Delta(i;\mu,t).
\end{equation}
The LSI describes the local inhomogeneity of a water molecule. A high local tetrahedral
order corresponds to a high value of $I$. To characterize the local order, we thus calculate the
index $\bar{I}$, averaged over all molecules and over time.  For instance, one finds an average 
$\bar{I}=0.25$ \AA$^2$ for the ice phase.\cite{palmer_Nature2014} Instead, when the disorder 
prevails, the $\bar{I}\approx 0$.

\begin{figure}
\centering
\includegraphics[width=7.5cm]{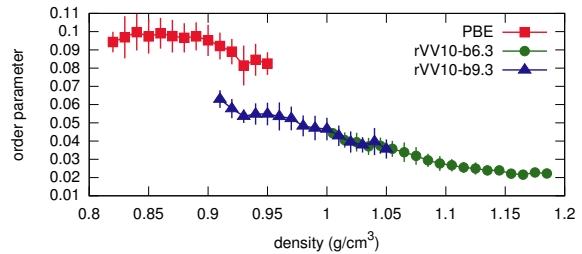}%
\caption{Average value of the local structure index (LSI) $I$ (Eq.\ \ref{eq:lsi}) as a
         function of the instantaneous density, as obtained from PBE, rVV10-b6.3, and 
         rVV10-b9.3 simulations. Vertical lines indicate the statistical error.}
 	\label{fig:lsi}
\end{figure}

Figure \ref{fig:lsi} shows the behavior of the LSI as a function of the instantaneous 
density in our molecular dynamics simulations. Results obtained with different functionals
are superposed. The PBE functional gives the lowest densities and the most ordered
local structures with $\bar{I}=0.10$ \AA$^2$. The rVV10 simulations give  
$\bar{I}=0.03$ \AA$^2$\ and $\bar{I}=0.05$ \AA$^2$\ for $b=6.3$ and $b=9.3$, respectively. 
Interestingly, the two rVV10 simulations yield local structure indices which superpose continously 
as a function of density, suggesting that the $I$ vs.\ density behaviour is not sensitive to 
the parameter $b$ within the class of rVV10 functionals. Furthermore, it is worthwhile
to point that the trend of LSI vs.\ density observed in this work is qualitatively 
consistent with the values of LSI found in simulations of high and low density
phases of liquid water in supercooled thermodynamic conditions.\cite{palmer_Nature2014}

\subsection{Hydrogen bond network}

It is customary to attribute the origin of the change in structural and dynamical
properties of liquid water caused by variations of external thermodynamic conditions
to the breaking or formation of hydrogen bonds and consequently to the rearrangement
of the molecules in the network.\cite{mills_JPC1973,krynicki_Farad1978,schwegler_PRL2000,todorova_JPCB2006}
The average number of hydrogen bonds is not a quantity which is directly measurable. 
However, a value of 3.58 per molecule has been inferred from experiments at 
ambient condition.\cite{soper_JCP1997}
In order to identify a hydrogen bond, we adopt a purely geometrical criterion 
which is commonly used in the literature.\cite{luzar_PRL1996,soper_JCP1997,schwegler_PRL2000}
We consider that two water molecules are hydrogen bonded when their oxygen atoms 
are separated by at most 3.5 \AA\ and simultaneously a hydrogen atom is located 
between them in such a way that the angle O-H-O is greater than 140$^{\circ}$.
Based on this criterion, we calculate the average number of hydrogen bonds per 
molecule in our molecular dynamics simulations, finding 3.73, 3.55, and 3.59, 
for PBE, rVV10-b6.3, and rVV10-b9.3 functionals, respectively. 
In particular, the value obtained for rVV10-b9.3 (3.59) is in very good agreement 
with the experimental estimate (3.58). 
Similar values have been obtained in previous simulations at fixed density in which 
van der Waals interactions were not taken into account.\cite{schwegler_PRL2000,todorova_JPCB2006} 
These results suggest that the average number of hydrogen bonds only undergoes 
minor variations, while the overall structural properties can be quite different.


\begin{table}
\caption{Distribution of water molecules with a given number of hydrogen bonds. 
          The average number of hydrogen bonds per water molecule is given in 
          the last column. The experimental estimate is reported for
          comparison.\cite{soper_JCP1997}}
\begin{ruledtabular}
\begin{tabular}{lcccccc}
   & \multicolumn{6}{c}{Number of hydrogen bonds} \\
   \cline{2-7}
   &  1 &  2 &  3 &  4 &  5 & average \\
   \hline
   PBE    &  0\% &  4\% & 20\% & 75\% &  1\% & 3.73 \\
   rVV10-b6.3  &  1\% &  9\% & 31\% & 52\% &  7\% & 3.55 \\
   rVV10-b9.3  &  1\% &  8\% & 28\% & 57\% &  6\% & 3.59 \\
   Expt.\cite{soper_JCP1997}  & -- & -- & -- & -- & -- & 3.58 
\end{tabular}
\end{ruledtabular}
\label{tab:hbond}
\end{table}

\begin{figure}
\centering
\includegraphics[width=6.5cm]{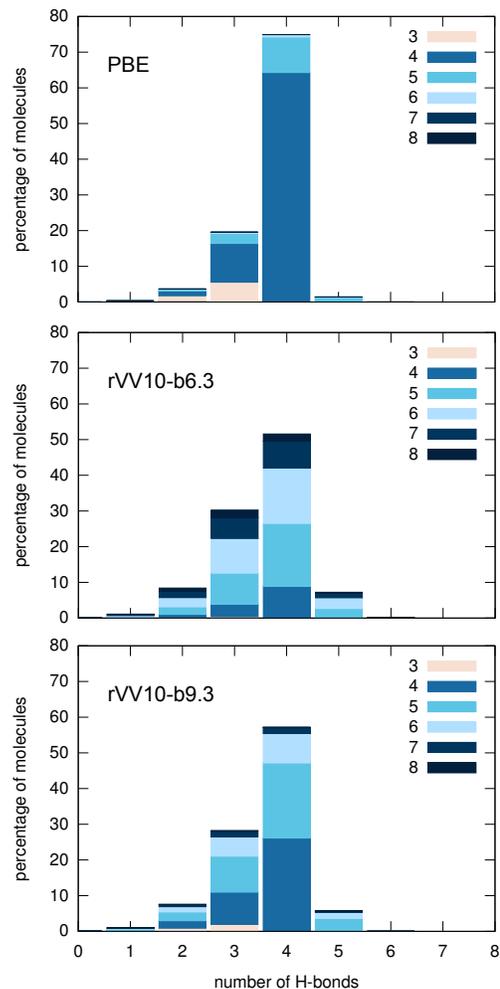}%
\caption{Distribution of water molecules with a given number of hydrogen bonds. 
         The finer subdivisions within each bar, indicated by a color code, 
         illustrate the associated distribution of coordination numbers.
         Since the same cutoff distance of 3.5 \AA\ is used for defining 
         the coordination number and the number of hydrogen bonds, 
         the former cannot be smaller than the latter.  
        }
         
 	\label{fig:hbond}
\end{figure}

In Table \ref{tab:hbond}, the percentage of molecules with a given number of 
hydrogen bonds is reported. In the PBE simulation, a large majority of molecules 
show four hydrogen bonds. Our result is in overall good agreement with a previous 
PBE study.\cite{todorova_JPCB2006} When van der Waals interactions are explicitly 
accounted for, the percentage of water molecules with either less or more than 
four hydrogen bonds increases. In particular, it can be seen from Table \ref{tab:hbond}
that there are no significant differences between the percentage distributions 
achieved with rVV10-b6.3 and rVV10-b9.3. However, we here show that 
this structural descriptor overlooks important structural differences which
relate to the occurrence of non-hydrogen bonded water molecules in the first
coordination shell.   

To highlight these structural differences, we illustrate the values given in 
Table \ref{tab:hbond} through a histogrammatic diagram in Fig.\ \ref{fig:hbond}. 
Each bar, corresponding to the percentage of water molecules with 
a given number of hydrogen bonds, is now further analyzed according to the
coordination number. The coordination number corresponds to the sum of all
water molecules within the first coordination shell, either hydrogen-bonded or not.
The first coordination shell is here defined by a cutoff distance of 3.5 \AA,
consistent with the adopted definition of hydrogen bond.\cite{luzar_PRL1996,schwegler_PRL2000,palmer_Nature2014}
Let us focus on the most frequent situation corresponding to molecules with four hydrogen
bonds. In the PBE simulation, most of these molecules are fourfold coordinated 
and only a small fraction of them is fivefold coordinated. The decomposition 
is very different for the rVV10 simulations, with the appearance of significant 
fractions of fivefold, sixfold, and sevenfold coordinated molecules. In particular,
while the average number of molecules with four hydrogen bonds are similar in rVV10-b9.3 
and rVV10-b6.3 (cf.\ Table \ref{tab:hbond}), the finer subdivision clearly shows that
the amount of highly coordinated molecules increases when going from rVV10-b9.3 to 
rVV10-b6.3.

We conclude our analysis of the hydrogen-bond network by focusing on the 
hydrogen bonds formed by the ``interstitial'' fifth water molecule within 
the first coordination shell of a central molecule. We focus on the most 
common situation in which a fifth molecule occurs in the coordination shell 
of a central molecule which is already engaged in four hydrogen bonds with the
first four molecules. In such a situation, the fifth molecule could form a 
varying number of hydrogen bonds with the water molecules in the first 
coordination shell of the central one: either none, one, or two. 
In Table \ref{tab:fifth-mol}, we report the corresponding distributions for  
our simulations.  
In the PBE simulation, the fifth molecule generally forms at least one of such 
hydrogen bonds with a large probability (76\%). At variance, in the rVV10
simulations, the most likely bonding corresponds to a fifth molecule without
any of such hydrogen bonds. These results further confirm that the density 
increase in rVV10 simulations is not driven by the hydrogen-bond energy 
but rather to the enhanced stability of more compact water arrangements, 
as expected for the non-directional nature of the van der Waals interactions.

\begin{table}
 \caption{Distribution of number of hydrogen bonds formed by a fifth molecule 
          in the coordination shell of a central molecule which is already engaged 
          in four hydrogen bonds with the first four molecules. }
 \label{tab:fifth-mol}
 \begin{ruledtabular}
  \begin{tabular}{lccc}
   & \multicolumn{3}{c}{Number of hydrogen bonds} \\
   \cline{2-4}
   &  0 &  1 &  2 \\
   \hline
   PBE    & 24\% & 55\% & 21\% \\
   rVV10-b6.3  & 60\% & 37\% &  3\% \\
   rVV10-b9.3  & 50\% & 44\% &  6\% \\
  \end{tabular}
 \end{ruledtabular}
 
\end{table}

\subsection{Diffusion coefficient}

To investigate the effect of van der Waals interactions on the mobility of water
molecules, we focus on the self-diffusion coefficient.
For a correct evaluation of the water self-diffusion coefficient, we perform 
a molecular dynamics simulation in the microcanonical ensemble (NVE).
\footnote{Since a NVE simulation does not require the calculation of stress
tensor, an energy cutoff of 70 Ry is sufficient for achieving convergence.}
We use the rVV10-b9.3 functional and a simulation cell at a density of 1 g/cm$^3$,
which closely corresponds to the equilibrium density of liquid water for this functional.
The average temperature in the simulation does not show any drift and averages to the
value of 350 K. We collect data on an equilibrated simulation run of 25 ps.

The mean square displacement vs.\ time is determined as an average over all molecules 
and over trajectories with initial times separated by 100 fs. 
By virtue of Einstein's relation, the calculated diffusion coefficient is derived 
from the slope of the mean square displacement and amounts to 
$D^\textrm{sim}=1.5 \cdot 10^{-5}$ cm$^2$/s. 

\begin{table}
\caption{Self-diffusion coefficient $D^\textrm{sim}$ of liquid water as obtained in our NVE simulation
         based on the rVV10-b9.3 functional. The result of a previous PBE simulation performed 
         at same density and temperature is also shown (Ref.\ \onlinecite{todorova_JPCB2006}). 
         The diffusion coefficient obtained in the simulations are compared to reference 
         values $D^\textrm{ref}$, which represent the experimental values for the cell size 
         used in the simulation. For reference, we also give experimental values at 300 and 
         350 K as derived from Ref.\ \citenum{mills_JPC1973}.}
\begin{ruledtabular}
\begin{tabular}{lccc}
     & $T$ (K) & $D^\textrm{sim}$ (cm$^2$/s)   & $D^\textrm{ref}$  (cm$^2$/s)\\
  \hline
  rVV10-b9.3                                   & 350 & $1.5  \cdot 10^{-5}$ & $4.3 \cdot 10^{-5}$  \\
  PBE (Ref.\ \onlinecite{todorova_JPCB2006})   & 350 & $0.47 \cdot 10^{-5}$ & $3.8 \cdot 10^{-5}$ \\
\hline
\\
&&& $D^\textrm{expt}$  (cm$^2$/s)\\ 
\hline
  Expt.(extrapolated)                          & 350 &      & $6.2 \cdot 10^{-5}$ \\
  Expt.                                        & 300 &      & $2.4 \cdot 10^{-5}$ 
 \end{tabular}
 \end{ruledtabular}
\label{tab:msd}
\end{table}

The comparison of the calculated diffusion coefficient with experiment requires
some care. Since the diffusion coefficient shows a strong dependence on
temperature,\cite{mills_JPC1973} it is necessary to compare theoretical and
experimental results achieved at the same temperature $T$.
Based on experimental data in Ref.\ \citenum{mills_JPC1973}, we obtain an 
extrapolated experimental value of $D^\textrm{expt.}=6.2 \cdot 10^{-5}$ cm$^2$/s
at 350 K, corresponding to the temperature in our simulation. 

The comparison between theory and experiment is further affected by 
finite-size effects. To estimate these effects, we assume that the diffusion
coefficient in the periodic cell underestimates the value corresponding to the
infinite-size limit by: \cite{kremer_JCP1993,yeh_JPCB2004}
\begin{equation}
\Delta D = \frac{k_\textrm{B}T\xi}{6\pi\eta L}
\end{equation}
where $k_\textrm{B}$ is the Boltzmann constant, $\xi=2.84$ a constant,
$\eta$ the viscosity, and $L$ the side of the periodic cell.
To take these finite-size effects into account we compare the calculated diffusion coefficient $D^\textrm{sim}$
with a reference value $D^\textrm{ref}$, which represent the experimental value
for the temperature and cell size used in the simulation. 
For this purpose, we take for $\eta$ the experimental value of the viscosity at 350 K
($\eta^\textrm{expt}=3.12\cdot 10^{-4}$ Pa$\cdot$s) as extrapolated from the data 
in Ref.\ \citenum{viscosity}, and for $L$ the cubic cell side corresponding
to 64 molecules at the experimental density ($L=12.4$ \AA). For the conditions
in our simulation, we thus obtain $\Delta D\approx 1.9\cdot 10^{-5}$ cm$^2$/s 
and $D^\textrm{ref}= 4.3 \cdot 10^{-5}$ cm$^2$/s. In Table \ref{tab:msd}, the diffusion 
coefficient $D^\textrm{sim}$ obtained from the molecular dynamics trajectory 
is directly compared with this reference value $D^\textrm{ref}$.

The comparison in Table \ref{tab:msd} indicates that the diffusion coefficient
obtained with the rVV10-b9.3 functional at 350 K ($D^\textrm{sim}= 1.5\cdot 10^{-5}$ cm$^2$/s) 
underestimates the reference value at the same temperature by a factor of about three.
This discrepancy should be attributed to the cumulative effect resulting
from the residual shortcomings of the adopted functional and the neglect of the 
quantum motion of the nuclei. However, the agreement is much better in case 
the calculated diffusion coefficient at 350 K is compared with a reference value
derived from the experimental diffusion coefficient at 300 K 
($1.8 \cdot 10^{-5}$ cm$^2$/s).

It is also instructive to compare the diffusion coefficients obtained with 
the rVV10-b9.3 functional with a previous PBE simulation at the same density and 
temperature.\cite{todorova_JPCB2006}
The present simulation enhances the diffusion coefficient approximately 
by a factor of three, yielding an improved agreement with experiment. 
Indeed, the diffusion coefficient $D^\textrm{sim}$ obtained in the present 
rVV10-b9.3 simulation corresponds to 35\%\ of the value of $D^\textrm{ref}$,
whereas the $D^\textrm{sim}$ in the PBE simulation only reaches 12\%\
of the respective reference value. Hence, the use of the present nonlocal 
van der Waals functional leads to an improved description of the water 
molecule mobility with respect to that achieved with the semilocal 
PBE functional.

\section{Conclusions}
\label{sec:conclusions}

We investigated the effect of explicitly including van der Waals interactions
in \textit{ab initio} molecular dynamics simulations of liquid water. 
Among the several available density functionals for treating van der Waals 
interactions, we adopted the rVV10 functional, which corresponds to a revised 
form\cite{sabatini_PRB2013} of the formulation proposed by Vydrov and Van Voorhis.\cite{vydrov_JCP2010}
This choice was motivated by the fact that the rVV10 functional allows for an evaluation of the total 
energy and the atomic forces at the same computational cost of  standard semilocal 
functionals. Furthermore, once the phenomenological parameters in the rVV10 functional 
are set, the same theoretical scheme can be carried over to study aqueous solutions.

We first checked the accuracy of the nonlocal rVV10 functional in describing the water dimer 
and ice I$_\textrm{h}$, finding energetic and structural properties in good 
agreement with both experimental and previous theoretical results. For these systems,  
the hydrogen-bond energy obtained with the rVV10 functional is in good agreement
with both the experimental and the PBE result.

We then performed molecular dynamics simulations of liquid water in the 
$NpH$ statistical ensemble. In particular, we compared results obtained through 
simulations using PBE and rVV10 functionals. With the semilocal PBE functional,
we obtained an equilibrium density of 0.87 g/cm$^3$, significantly lower than the
experimental density of 1 g/cm$^3$, but in very good agreement with 
previous $NpT$ simulations based on the same functional.\cite{schmidt_JPCB2009}
At variance, the rVV10 functional, in which the parameter $b$ is set to its original value 
(6.3, Ref.\ \onlinecite{sabatini_PRB2013}), yields a density of 1.1 g/cm$^3$.
We found that the equilibrium density could be varied by modifying the
parameter $b$. For $b=9.3$, we obtained an equilibrium density of 0.99 g/cm$^3$,
very close to the experimental value. 
Interestingly, we noticed that the use of rVV10-b9.3 also improved the local structure 
and the mobility of the water molecules. Indeed, the oxygen-oxygen radial distribution 
function in rVV10-b9.3 is in much better agreement with experiment than PBE.
Furthermore, the average number of hydrogen bonds per molecule achieved with the
rVV10-b9.3 functional (3.59) shows a closer agreement with the estimate derived
from experimental data (3.58), than the corresponding numbers obtained with PBE
(3.73) or rVV10-b6.3 (3.55).

The higher density achieved for liquid water with the nonlocal van der Waals 
functional indicates that the dispersion forces promote the role of 
non-directional interactions with respect to the semilocal functional in 
which the directionality of the hydrogen bonds dominates. 
The observed structural variations in liquid water thus arise from a different 
balance between non-directional and directional interactions. 

In conclusion, our work demonstrates that the explicit treatment of van der Waals 
interactions through the rVV10 formulation considerably improves both 
the structural and the dynamical properties of liquid water with respect 
to the Perdew-Burke-Ernzerhof functional. This improvement is 
achieved without any significant computational overhead. 
Hence, the present theoretical framework should generally be preferred over 
the use of the Perdew-Burke-Ernzerhof functional in any application involving 
liquid water.

\acknowledgments
This work has been performed in the context of the National Center of Competence
in Research (NCCR) ``Materials' Revolution: Computational Design and Discovery
of Novel Materials (MARVEL)'' of the Swiss National Science Foundation. We used 
computational resources of CSCS and CSEA-EPFL. Stefano de Gironcoli gratefully 
acknowledges hospitality and support by the CECAM headquarters in Lausanne.

\end{document}